\documentclass[a4paper,11pt]{article}
\pdfoutput=1 

\usepackage{jheppub} 
\usepackage[T1]{fontenc} 

\usepackage{amssymb}

\usepackage[utf8]{inputenc} 
\usepackage{graphicx}

\usepackage{braket}
\usepackage[vcentermath]{youngtab}

 \usepackage{fancyhdr}

\usepackage{mathrsfs}
\usepackage[T1]{fontenc}
\usepackage{setspace}
\usepackage{amsfonts}
\usepackage{amssymb}
\usepackage{amsmath}
\usepackage{epsfig}
\usepackage{latexsym}
\usepackage{color}
\usepackage{nicefrac}
 \usepackage{slashed}
 \usepackage{multirow}
 \usepackage{comment}
 \usepackage{soul}
 \usepackage{hyperref}
\usepackage{slashed}
\usepackage{simpler-wick}
\usepackage{array}
\usepackage{ textcomp }
\usepackage{tabu}
\usepackage{tcolorbox}
\usepackage{subcaption}
\newcommand{\ea}[1]{
\begin{align}
#1
\end{align}
}

\newcommand{\beq}{\begin{eqnarray}}
\newcommand{\eeq}{\end{eqnarray}}

\newcommand{\p}{{\cal P}\exp}

\newcommand{\bmp}{\noindent\begin{minipage}{16cm}}
\newcommand{\emp}{\end{minipage}\vskip 7mm} 

\renewcommand \l  {\lambda}

\newcommand*{\del}{\mathop{\mathrm{{}\partial}}\mathopen{}}

\newcommand{\ee}{\end{equation}}
\newcommand{\be}{\begin{equation}}
\newcommand{\bea}{\begin{align}}
\newcommand{\eea}{\end{alig}}

\def\lsim{\mathrel{\rlap{\lower4pt\hbox{\hskip1pt$\sim$}}
    \raise1pt\hbox{$<$}}}                
\def\gsim{\mathrel{\rlap{\lower4pt\hbox{\hskip1pt$\sim$}}
    \raise1pt\hbox{$>$}}}                
    
    \def\be{\begin{equation}}
\def\ee{\end{equation}}
\def\ba{\begin{eqnarray}}
\def\ea{\end{eqnarray}}

\def\del{\partial}

\def\a{\alpha}

\def\e{\epsilon}

\def\p{\pi}

\def\l{\lambda}

\def\IR{\relax{\rm I\kern-.18em R}}

\def\IR{\relax{\rm I\kern-.18em R}}
\def\IL{\relax{\rm I\kern-.18em L}}

\def\inv{^{\raise.15ex\hbox{${\scriptscriptstyle -}$}\kern-.05em 1}}

\title{\boldmath Boundary Symmetry Breaking in CFT and the String Order Parameter}

\author[a]{Riccarda Bonsignori,}
\author[b]{Luca Capizzi,}
\author[a]{Pantelis Panopoulos}

\affiliation[a]{Rudjer Boskovic Institute, Division of Theoretical Physics, Bijeni\v cka 54, 10000 Zagreb, Croatia}

\affiliation[b]{SISSA and INFN Sezione di Trieste, via Bonomea 265, I-34136 Trieste, Italy}

\emailAdd{rbonsign@irb.hr}
\emailAdd{lcapizzi@sissa.it}
\emailAdd{Pantelis.Panopoulos@irb.hr}

\abstract
{
We consider the ground state of a one-dimensional critical quantum system carrying a global symmetry in the bulk, which is explicitly broken by its boundary conditions. We probe the system via a string-order parameter, showing how it detects the symmetry breaking pattern. We give a precise characterization of the mechanism depicted above in Boundary CFT, and we find a general logarithmic scaling for the order parameter. As a first example we analyze the breaking of a $U(1)$ symmetry for complex free theories induced by a boundary pairing term. Moreover, we give predictions for the breaking of $U(N)$ in free theories, arising from a boundary mixing. We test our predictions with numerical calculations for some lattice realizations of free fermionic system with boundary symmetry breaking, finding a good agreement.  
}


\begin{document}

\maketitle
\flushbottom

\section{Introduction}

Symmetry is a cornerstone of modern physics. Its importance emerges in all branches of physics among which the Condensed Matter Theory \cite{sh-17, a-04} (ferromagnetism, superconductivity, superfluidity) and High-Energy Physics \cite{w-95} (formulation of the Standard Model,  Quantum gravity etc.). There are many cases where symmetry is responsible for phase transitions in statistical systems \cite{zj-05,s-11}. In particular, behind plenty of classical and quantum systems showing a (second order) phase transition there is an underlying symmetry that can be preserved or spontaneously broken: this mechanism has been a guideline for the characterization of the possible phases of the system.

As an illustrative example, we consider a toy model for ferromagnetism, namely the classical Ising model enjoying global symmetries. It has been shown that in $d\geq 2$ dimensions, the model undergoes a phase transition at a critical value of the temperature $T=T_c$. Moreover, for $T>T_c$ the $\mathbb{Z}_2$ symmetry of the model is unbroken, and the system has a paramagnetic behavior, while for $T<T_c$ the system undergoes to a Spontaneous Symmetry Breaking (SSB) characterizing the ferromagnetic phase. The breaking of the symmetry can be spotted via the probing of the magnetization, described in terms of a local field $\sigma(x)$. In particular, its expectation value $\langle \sigma(x)\rangle$ vanishes at $T>T_c$ and it gets a finite value for $T<T_c$. The critical point $T=T_c$ deserves special attention. Indeed, while at this point the symmetry remains unbroken and thus $\langle \sigma(x)\rangle=0$, an additional application of a magnetic field in this critical temperature can have non-trivial effects even far from that point, due to the slow (algebraic) decay of the correlation functions, a distinct 
 trait of the critical phase.

Historically, the importance of localized perturbations for critical systems has not been taken for granted, until the discovery of the Kondo effect \cite{k-64,a-95}, which is associated with anomalous transport properties of low-temperature metals in the presence of impurities. Despite the progresses in this field, both experimentally and theoretically (e.g. Boundary Conformal Field Theory (BCFT) formulation \cite{c-84,c-89,c-04}), a comprehensive theory of the symmetry and its breaking pattern seems to remain an open problem for such systems. The main motivation behind this work is  to address more systematically this lack of approach.

We are interested in the ground-state $|\Omega\rangle$ of the following Hamiltonian
\be
\mathcal{H} = \mathcal{H}_0+\mathcal{H}_1\,\,.
\ee
 The $\mathcal{H}_0$ is a bulk term involving short-range interactions, which we assume to be critical, and invariant under the action of a group $G$ associated with a global symmetry. In turn, the $\mathcal{H}_1$ term is a   perturbation localized in space (say defect/impurity) which explicitly spoils the $G$-invariance. More precisely, we consider a unitary representation of $G$, which associates to any $g\in G$ a unitary operator $\hat{g}$ of the Hilbert space, and we require that
\be\label{Hor}
[\mathcal{H}_0,\hat{g}] = 0, \qquad [\mathcal{H}_1,\hat{g}] \neq 0, \qquad \forall g \in G.
\ee
Our main goal is to understand whether and how the ground state $\ket{\Omega}$ of \eqref{Hor} breaks the global $G$-invariance. For this
purpose, we propose
\be
\bra{\Omega}\hat{g}\ket{\Omega},
\ee
regarded as a function of G, to be a good non-local order parameter.
For one-dimensional systems, $\hat{g}$ can be naively regarded as a string operator inserted along the whole system, and for this reason, we call it \emph{string order parameter} (following the terminology of \cite{kt-92}). We anticipate that, rather generically, the unbroken symmetry group is identified by
\be\label{eq:unbrok_group}
H = \{ g \in G | \  |\bra{\Omega}\hat{g}\ket{\Omega}|=1 \},
\ee
while for the other elements of the group  one has  $|\bra{\Omega}\hat{g}\ket{\Omega}|<1$. Moreover, for a critical system defined on a finite-size region $[0,L]$, we find that in the presence of scale-invariant symmetry-breaking boundary terms, the (log of the) order parameter shows the following logarithmic growth
\be\label{eq:log_growth}
-\log |\bra{\Omega}\hat{g}\ket{\Omega}| \sim \log L,
\ee
in the limit of large $L$. In particular the quantity 
\be
\underset{L\rightarrow \infty}{\lim}-\frac{\log |\bra{\Omega}\hat{g}\ket{\Omega}|}{\log L}, \quad g \in G
\ee
is a universal continuous, but in general not smooth function of the group $G$, which vanishes precisely for $g \in H$, with $H$ being the unbroken subgroup.

We provide an accurate description of this mechanism in the context of BCFT, exploiting the power of conformal symmetry. Therefore, we compute explicitly the string order parameter for a class of free fermionic and bosonic theories. In the first case, we consider a free Dirac fermion on $[0,L]$ and we insert a pairing term at one boundary point, studying the symmetry breaking pattern $U(1)\rightarrow \mathbb{Z}_2$. We extend our analysis by taking $N$ copies of uncoupled the Dirac fields  in the bulk, but  coupled  at one boundary point via a scattering matrix $S$. In this case, we find a novel non-trivial breaking of the $U(N)$ symmetry, strongly related to the symmetries of $S$. By  repeating  the same approach we study the string order parameter for free complex massless bosons under $U(1)$ and  under $U(N)$ symmetry seperately as well. Finally, we present two possible realizations of the boundary symmetry breaking mechanism for free fermionic systems on the lattice. For one of them, we also perform the numerical calculation of the string order parameter to test our analytical prediction.

Our manuscript is organized as follows. In Sec. \ref{sec:Def} we provide some general definitions and  give a BCFT description of the string order parameter. In Sec. \ref{sec:Fermions} we analyze in detail the free fermions. In Sec. \ref{sec:Bosons} we repeat the same analysis for free bosons. In Sec. \ref{sec:Lattice} we consider the lattice counterpart for fermions and present the numerical results. Finally, in Sec. \ref{sec:Conclusions} we gather our results and discuss some possible future directions.

\section{Definitions and Techniques}\label{sec:Def}

The purpose of this section is two-fold. First, once the non-local order parameter is introduced, we explain how and why it detects symmetry-breaking, providing some properties which are independent of the detail of the systems. Then, we specify the treatment to 1+1 BCFT with symmetry breaking terms at the boundaries, and give a general derivation of the logarithmic growth in Eq. \eqref{eq:log_growth}. At this point, we will not specialize to any specific theory, and we derive the first results employing only conformal symmetry  and describing the boundary conditions (BCs) as boundary states, via a space-time duality.\\

\subsection{String order parameter}

Let us first review what is a symmetry in a quantum system. Given a Hilbert space $\mathscr{H}$, a unitary representation of a group $G$ is a linear map
\be
\begin{split}
&G\rightarrow  GL(\mathscr{H})\,,\qquad g \rightarrow \hat{g}
\end{split}
\ee
from the group onto the unitary operators of $\mathscr{H}$. One requires that the map is homomorphic, namely
\be
\hat{(g_1g_2)} = \hat{g_1}\hat{g_2}\,\,.
\ee
Without loss of generality, here we assume that the map is injective, so that distinct elements of the group are represented by distinct operators. 
We now consider a (normalized) state $\ket{\Omega}\in \mathscr{H}$. We say that $\ket{\Omega}$ is symmetric (invariant) under $G$ iff
\be\label{eq:g_inv}
\hat{g}\ket{\Omega} = e^{i\phi(g)}\ket{\Omega}, \quad \forall g\in G
\ee
with $e^{i\phi(g)}$ being a $g$-dependent phase factor. The requirement above severely constrains the expectation values of the observables, which is the main reason why the order parameters are useful to detect the breaking of symmetry.

As a first example, we consider a quantum system carrying a representation of $\mathbb{Z}_2$. We denote by $\{1,\tau\}$ the generators of the group, and consider an observable $\sigma$, say the magnetization, odd under $\mathbb{Z}_2$
\be
\hat{\tau} \sigma \hat{\tau}^{-1} = -\sigma,
\ee
which plays the role of the order parameter. Whenever a state $\ket{\Omega}$, say the (a) ground state, is invariant under $\mathbb{Z}_2$ one safely concludes that
\be
\bra{\Omega}\sigma\ket{\Omega} = \bra{\Omega}\hat{\tau}^{-1}\sigma\hat{\tau}\ket{\Omega} = -\bra{\Omega}\sigma\ket{\Omega},
\ee
which clearly implies that $\bra{\Omega}\sigma\ket{\Omega} =0$. This means that, whenever $\bra{\Omega}\sigma\ket{\Omega} \neq 0$ one can be sure that the state $\ket{\Omega}$ is not $\mathbb{Z}_2$ invariant. Unfortunately, in principle, the converse is not true. Indeed, there is no reason why $\bra{\Omega}\sigma\ket{\Omega} =0$ should imply a $\mathbb{Z}_2$ symmetry for $\ket{\Omega}$. This is somehow the main disadvantage behind the usage of the usual order parameters.

In contrast, as we will show below, if one considers $\hat{g}$ itself as an order parameter, one can unambiguously understand if $\ket{\Omega}$ is symmetric. A first immediate observation is that, if $\ket{\Omega}$ is symmetric under $g \in G$ (see Eq. \eqref{eq:g_inv}), then
\be
|\bra{\Omega}\hat{g}\ket{\Omega}|=|\bra{\Omega}e^{i\phi(g)}\ket{\Omega}| =1.
\ee
Less trivially, one can show the converse, namely that $|\bra{\Omega}\hat{g}\ket{\Omega}| = 1$ implies that the symmetry has to be unbroken. To prove it, we employ the Cauchy-Schwarz inequality \cite{CSI}, which tells us
\be
|\bra{\Omega}\hat{g}\ket{\Omega}| \leq |\bra{\Omega}\hat{g}^\dagger \hat{g}\ket{\Omega}|\cdot |\braket{\Omega|\Omega}|=1,
\ee
where the inequality is strict unless $\ket{\Omega}$ and $\hat{g}\ket{\Omega}$ are proportional, which is exactly the notion of symmetry in Eq. \eqref{eq:g_inv}. To summarize, so far we have that
\be
|\bra{\Omega}\hat{g}\ket{\Omega}| =1 \quad\text{ if and only if } \quad\hat{g}\ket{\Omega} = e^{i\phi(g)}\ket{\Omega}.
\ee
This property suggests a way to characterize the subgroup $H \subseteq G$ which leaves $\ket{\Omega}$ invariant as Eq. \eqref{eq:unbrok_group}. We would like to stress explicitly that, in their simplicity, the last conclusions are very general and apply to both abelian and non-abelian symmetries. Notice that up to this point  our discussion is general, since we did not specify whether the symmetry breaking pattern $G\rightarrow H$, associated with the state $\ket{\Omega}$, arises from an explicit or spontaneous symmetry breaking. For the reasons depicted above, the investigation of $|\bra{\Omega}\hat{g}\ket{\Omega}|$, the non-local (string) order parameter, is the main goal of this work.

\subsection{BCFT description}\label{subsec:BCFT}

\begin{figure}[t]
\centering
	\includegraphics[width=0.9\linewidth]{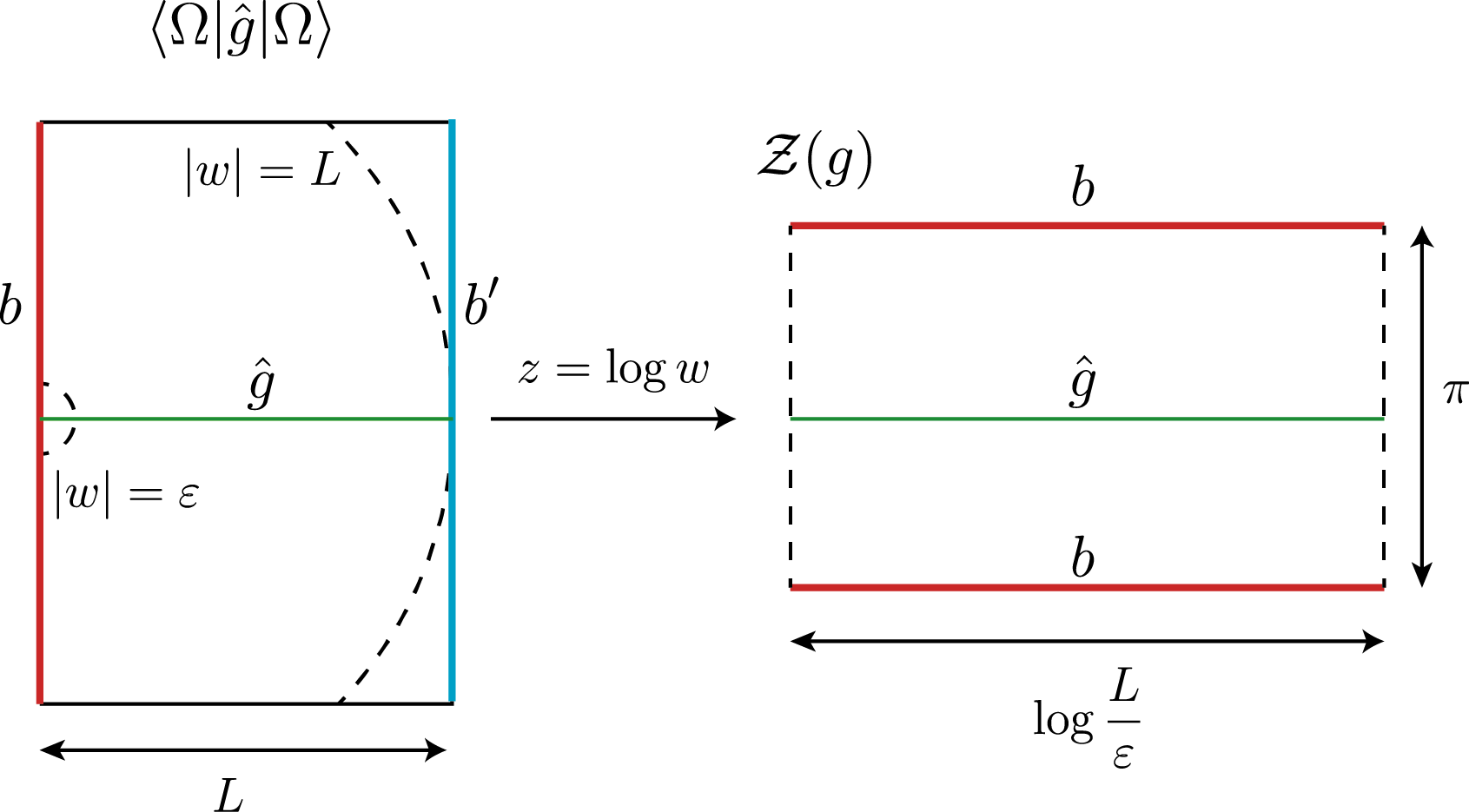}
    \caption{The expectation value of the string order parameter. We represent the original strip geometry (coordinate $w$) and the rectangular one (coordinate $z$). The red/blue lines correspond to the BCs of type $b,b'$ respectively. The insertion of the symmetry operator $\hat{g}$ is a green line.
    }
    \label{fig:Charged_pfun}
\end{figure}

Let us proceed to the description of the string order parameter in the framework of BCFT
\cite{DiFrancesco-97,ss-08,bb-15,cmc-22,cmc2-22}. 
We consider the ground state $\ket{\Omega}$ of a one-dimensional critical quantum system on a finite size geometry, namely the interval $[0,L]$. We assume that the bulk is described by a CFT, and we impose conformal invariant BCs  at $x=0,L$ \cite{DiFrancesco-97,c-84}. Introducing the Euclidean time, one can describe the state $\ket{\Omega}$ as a strip geometry, parametrized by the complex coordinate $w$ satisfying
\be
\text{Re}(w) \in [0,L], \quad  \text{Im}(w) \in (-\infty,\infty).
\ee
In particular, $\text{Re}(w)$ represents the spatial position, and $\text{Im}(w)$ corresponds to the Euclidean time. In this picture, the expectation values of the observables in the state $\ket{\Omega}$ are described via the insertion of fields in the strip geometry depicted in Figure \ref{fig:Charged_pfun}.

We now assume that the theory has a global symmetry in the bulk, characterized by a representation of a group $G$, which may be eventually broken by the choice of the BCs. We are interested in the symmetry breaking pattern, strictly related to the evaluation of the string order parameter $\bra{\Omega}\hat{g}\ket{\Omega}$. Since the symmetry is global, the action of the group $G$ is nontrivial at any spatial point $x \in [0,L]$. For this reason, it is natural to represent pictorially $\hat{g}$ as a line operator extended over $\text{Im}(w)=0$. Then, $\bra{\Omega}\hat{g}\ket{\Omega}$ becomes a charged partition function given by the insertion of a charged line connecting $w=0$ and $w=L$ in the strip geometry.

The last key ingredient is the specification of the BCs. We denote by $b,b'$ the type of BCs at $x=0,L$ respectively. In the strip geometry, these boundary points become lines extended over the euclidean time, $\text{Re}(w)=0,L$ respectively, and we associate the labels $b,b'$ to each of these lines. We assume that $b'$, corresponding to $x=L$, preserves explicitly the symmetry, and we do not characterize it further. Instead, we focus on BCs at $x=0$ which breaks $G$ explicitly, and our goal is to identify the corresponding symmetry breaking pattern.

At this point, we employ conformal symmetry to relate the strip geometry described above to another geometry and  we do that to simplify the computation of the charged partition function. After a UV and IR regulation of the original geometry, keeping only the points $\varepsilon<|w|<L$, we apply the transformation \cite{hlw-94,cw-94,ct-16,ot-15}
\be
z= \log w.
\ee
The new geometry is the rectangle
\be
\text{Re}(z) \in (\log \varepsilon,\log L), \quad \text{Im}(z) \in [-\pi/2,\pi/2],
\ee
with BCs of type $b$ along $\text{Im}(z) = \pm \pi/2$, corresponding to boundary states. It is important to stress that the information about $b'$ is lost explicitly by the choice of IR regularization. However, this is not big deal, as we required that $b'$ is $G$ invariant. Indeed, on the physical ground, we do not expect any contribution to the string order parameter from the point $x=L$ (at least at leading order).

Putting these information together, we express the charged partition function as a transition element in Euclidean time between the boundary state $\ket{b}$, corresponding to $b$ (see Ref.\cite{DiFrancesco-97}), with itself. More precisely, we define
\be\label{eq:charged_pfun}
\mathcal{Z}(g) \equiv \bra{b}\exp( -\pi H)\hat{g}\ket{b},
\ee
with $H$ being the Hamiltonian in the rectangular geometry
\be
H = \frac{2\pi}{\log(L/\varepsilon)}(L_0+\bar{L}_0),
\ee
and $L_0,\bar{L}_0$ the Virasoro generators. Then, we express the expectation value of $\hat{g}$ as
\be\label{eq:pfunc_rel}
\bra{\Omega}\hat{g}\ket{\Omega} = \frac{\mathcal{Z}(g)}{\mathcal{Z}(1)},
\ee
where the denominator $\mathcal{Z}(1)$ arises only as a normalization constant (the uncharged partition of the rectangle). We represent the construction above in Fig. \ref{fig:Charged_pfun}, showing the insertion of $\hat{g}$ in the two geometries.
So far, the discussion is general and it applies to any BCFT carrying a global symmetry $G$. The precise evaluation of the charged partition function $\mathcal{Z}(g)$ requires the characterization of the boundary state $\ket{b}$. Still, in general, one can argue that the logarithm of the partition function (free energy) is extensive in the large-size limit. In particular, we have
\be
\log \mathcal{Z}(g) \propto \log \frac{L}{\varepsilon}, \qquad L/\varepsilon \rightarrow \infty
\ee
up to a proportionality constant depending on both $g \in G$ and the BCs $b$. Summarizing, we discover that the ratio 
\be
-\frac{\log |\bra{\Omega}\hat{g}\ket{\Omega}|}{\log L/\varepsilon}
\ee
is finite in the limit $L\rightarrow \infty$, where the conventional minus sign ensures  a positive value result. In the next section, we will carefully analyze that ratio for specific CFTs, relating its behavior to the symmetry breaking pattern.

\section{Dirac fermions}\label{sec:Fermions}

We now proceed with an application of the previous discussion to the system of free fermions. First, we discuss free fermions with $U(1)$ symmetry and then we continue with $U(N)$ generalization\footnote{For more details in the calculations of this section the reader may consult \cite{cmc-22}.}.

\subsection{$U(1)$-symmetry breaking due to a boundary pairing term}\label{sec:U1_ferm}

We consider the theory of massless Dirac fermions in a finite-size geometry, taking BCs that break explicitly the $U(1)$ symmetry. The model is described by two fields $\Psi$ and $\Psi^\dagger$, that correspond to particle and antiparticles. In radial quantization, one decomposes $\Psi$ in its left and right Laurent modes as follows
\be
\Psi(z) = \sum_{k\in \mathbb{Z}+1/2}\frac{\Psi_k}{z^{k+1/2}}, \qquad \bar{\Psi}(\bar{z}) = \sum_{k\in \mathbb{Z}+1/2}\frac{\bar{\Psi}_k}{\bar{z}^{k+1/2}},
\ee
where the Neveu-Schwarz (NS) sector has been considered. For $k>0$ the modes $\Psi_k,\bar{\Psi}_k$ destroy a left/right moving particle, while $\Psi_{-k},\bar{\Psi}_{-k}$ are creation operators. Similar considerations hold for $\Psi^\dagger$, the antiparticle field, and we refer to its left/right modes with $\Psi^\dagger_k,\bar{\Psi}^\dagger_k$. 

\begin{figure}[ht]
\centering
	\includegraphics[width=0.9\linewidth]{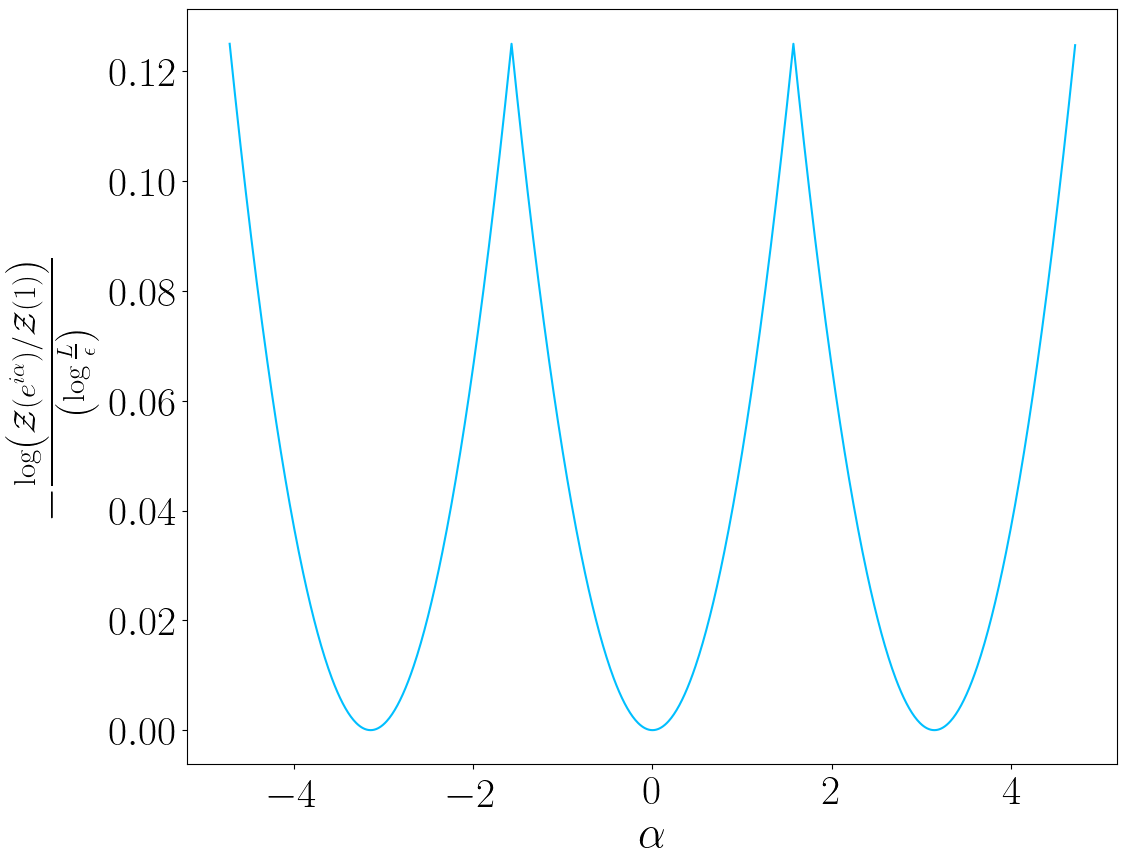}
    \caption{String order parameter as function of $\alpha$ for the Dirac fermion. The function is periodic for $\alpha \rightarrow \alpha + \pi$, and shows cusp singularities in correspondence of $\alpha= \frac{\pi}{2}+ k\pi, k\in \mathbb{Z}$. As expected, it vanishes at $\alpha= k\pi, k\in \mathbb{Z}$ and it signals the presence of an unbroken $\mathbb{Z}_2$ subgroup.
    }
    \label{fig:SOPfermions}
\end{figure}

The bulk  action in Euclidean space of the finite $[0,L]$ geometry reads 
\be
S_{\text{bulk}} = \int^L_0 dx\int d\tau  \left( \Psi^\dagger \bar{\partial} \Psi+\bar{\Psi}^\dagger \partial \bar{\Psi}\right).
\ee
 enjoying  a  $U(1)$ global symmetry 
\be
\Psi \rightarrow e^{i\alpha}\Psi, \quad \bar{\Psi} \rightarrow e^{i\alpha}\bar{\Psi}, \quad \Psi^\dagger \rightarrow e^{-i\alpha}\Psi^\dagger, \quad \bar{\Psi}^\dagger \rightarrow e^{-i\alpha}\bar{\Psi}^\dagger,
\ee
and it corresponds to the imbalance between particles and antiparticles. We want to break $U(1)$ explicitly through the BCs at $x=0$. We do so via the insertion of a pairing term at the boundary point, described by a boundary action
\be
S_{\text{boundary}} = \int d\tau \  \left(\bar{\Psi}(x=0,\tau)\Psi(x=0,\tau) + (\Psi \rightarrow \Psi^\dagger)\right).
\ee
The boundary term is not $U(1)$ invariant, since it transforms as $\Psi\bar{\Psi}\rightarrow e^{i2\alpha} \Psi\bar{\Psi}$. A residual $\mathbb{Z}_2$ is nevertheless preserved (associated with $\alpha=0,\pi$), and it describes the conservation of fermion parity. One thus expects that these BCs should induce an explicit symmetry breaking pattern
\be\label{U1breaking}
U(1) \rightarrow \mathbb{Z}_2,
\ee
on the ground state $\ket{\Omega}$. While these considerations are so far not rigorous, they can capture the key features of the system.
In the following, we aim to compute the string-order parameter via the BCFT techniques \eqref{subsec:BCFT} for the $U(1)$ symmetry. The first quantity we need is the boundary state $\ket{b}$ associated with the $U(1)$ breaking BCs. We consider a coherent state in which pairs of particles (and antiparticles) with opposite momenta are generated above the ground states. Its explicit expression is
\be\label{bstate}
\ket{b} = \prod_{k>0}\exp( i \Psi_{-k}\bar{\Psi}_{-k}  + (\Psi \rightarrow \Psi^\dagger))\ket{0},
\ee
with $\ket{0}$ being the vacuum of the theory. 
The generator of the symmetry is
\be\label{charge}
Q = \sum_{k>0}\Psi_{-k}\Psi_{k} + \bar{\Psi}_{-k}\bar{\Psi}_{k} - (\Psi \leftrightarrow \Psi^\dagger ),
\ee
and we parametrize a generic element of $U(1)$ as
\be\label{U1element}
\hat{g} = e^{i\alpha Q}, \quad \alpha \in [-\pi,\pi).
\ee
Finally, we remind the relation between the Virasoro modes and the fermionic modes
\be
L_0+\bar{L}_0 = \sum_{k>0} k(\Psi_{-k}\Psi_k + \bar{\Psi}_{-k}\bar{\Psi}_k+ (\Psi \leftrightarrow \Psi^\dagger)).
\ee
We proceed with the evaluation of the charged partition function $\mathcal{Z}(e^{i\alpha})$, associated with the $U(1)$ phase $e^{i\alpha}$. Putting the previous elements together into \eqref{eq:charged_pfun}, we get
\be\label{partition}
\mathcal{Z}(e^{i\alpha}) = \bra{b}e^{i\alpha Q}q^{L_0+\bar{L}_0}\ket{b},
\ee
where $q$ is defined, for later convenience, as
\be\label{eq:def_q}
q \equiv \exp \left( -\frac{2\pi^2}{\log (L/\varepsilon)}\right).
\ee
We further decompose $\mathcal{Z}(e^{i\alpha})$ as a product over the fermionic modes
\be
\begin{split}
\mathcal{Z}(e^{i\alpha}) = \prod_{k>0}\bra{0}\exp(-i\bar{\Psi}_{k} \Psi_{k})\exp(i\alpha \Psi_{-k}\Psi_{k}+i\alpha \bar{\Psi}_{-k}\bar{\Psi}_{k})\\
\times q^{k\Psi_{-k}\Psi_{k}+k\bar{\Psi}_{-k}\bar{\Psi}_{k}}\exp(i \Psi_{-k}\bar{\Psi}_{-k})\ket{0} \times (\Psi \rightarrow \Psi^\dagger).
\end{split}
\ee
The building block we need to proceed with is the contribution coming from the mode $k$, evaluated as
\be
\begin{split}
\bra{0}\exp(-i\bar{\Psi}_{k} \Psi_{k})\exp(i\alpha \Psi_{-k}\Psi_{k}+i\alpha \bar{\Psi}_{-k}\bar{\Psi}_{k}) q^{k\Psi_{-k}\Psi_{k}+k\bar{\Psi}_{-k}\bar{\Psi}_{k}}\exp(i \Psi_{-k}\bar{\Psi}_{-k})\ket{0}  =\\
\bra{0}\exp(-i\bar{\Psi}_{k} \Psi_{k})\exp(iq^{2k}e^{i2\alpha} \Psi_{-k}\bar{\Psi}_{-k})\ket{0}  = (1+q^{2k}e^{i2\alpha}),
\end{split}
\ee
where the commutation relations of the modes, together with the property $\Psi_{k}\ket{0} = \bar{\Psi}_{k}\ket{0} = 0$ (valid for $k>0$), have been employed. Putting it all together, we reach to
\be
\mathcal{Z}(e^{i\alpha}) = \prod_{k \in \mathbb{N}-1/2}(1+q^{2k}e^{i2\alpha})(1+q^{2k}e^{-i2\alpha}) = \prod^{\infty}_{m=1}(1+2\cos(2\alpha)q^{2m-1}+q^{4m-2}).
\ee
Before proceeding further, we observe that
\be
\mathcal{Z}(-1) = \mathcal{Z}(1),
\ee
that implies \eqref{eq:pfunc_rel}
\be
\bra{\Omega}e^{i\pi Q}\ket{\Omega} = \frac{\mathcal{Z}(-1)}{\mathcal{Z}(1)} = 1.
\ee
This means that the fermion parity, generated by $(-1)^Q$, is a symmetry of $\ket{\Omega}$, as expected. In addition, as we will show below, there are no additional symmetries, and $\mathbb{Z}_2$ is precisely the unbroken subgroup $H$ (see Eq. \eqref{eq:unbrok_group}).
We provide the explicit expression of $\mathcal{Z}(e^{i\alpha})$ in the limit of $L/\varepsilon \rightarrow 1$, converting the infinite product in an integral
\be
\log \mathcal{Z}(e^{i\alpha}) \simeq \int^\infty_0 dk \log(1+q^{2k} e^{i2\alpha})+\log(1+q^{2k} e^{-i2\alpha}) = \frac{1}{2\log q}\left(\text{Li}_2(-e^{i2\alpha})+\text{Li}_2(-e^{-i2\alpha})\right).
\ee
Using the properties of the dilogarithm function, and the definition of $q$ \eqref{eq:def_q} we reach the final expression of the order parameter
\be
\label{eq:sopfermions}
-\log \bra{\Omega}e^{i\alpha Q}\ket{\Omega} = - \log \frac{\mathcal{Z}(e^{i\alpha})}{\mathcal{Z}(1)}= \frac{\alpha^2}{2\pi^2}\log L/\varepsilon, \quad \alpha \in [-\pi/2,\pi/2],
\ee
whose values are periodic under $\alpha \rightarrow \alpha + \pi$. Since it vanishes for $e^{i\alpha}=\pm 1$, we identify the unbroken group \eqref{eq:unbrok_group} with $H = \{1,-1\}$, which is nothing but the fermionic parity. As a last remark, we notice the presence of a cusp singularity for $e^{i\alpha} =\pm i$, where the order parameter is continuous but not differentiable.
In Fig \ref{fig:SOPfermions}, we show the behavior of \eqref{eq:sopfermions} as a function of $\alpha$.

\subsection{$U(N)$-symmetry breaking  via boundary scattering matrix}\label{sec:UN_ferm}

Here, we consider $N$ Dirac fermions coupled together via a boundary scattering matrix. In particular, we take the bulk action
\be
S_{\text{bulk}} = \int^L_0 dx\int d\tau  \left( (\Psi^\dagger)^j \bar{\partial} \Psi^j+(\bar{\Psi}^\dagger)^j \partial \bar{\Psi}^j\right),
\ee
where the sum over $j=1,\dots,N$, the species index, is implicit. Then, we notice that the following $U(N)$ symmetry is present
\be\label{eq:UN_symmetry}
\Psi^j \rightarrow U_{j'j}\Psi^{j'}, \quad  \bar{\Psi}^j \rightarrow U_{j'j}\bar{\Psi}^{j'}, \quad  (\Psi^\dagger)^j \rightarrow \bar{U}_{j'j}(\Psi^\dagger)^{j'}, \quad 
(\bar{\Psi}^\dagger)^j \rightarrow \bar{U}_{j'j}(\bar{\Psi}^\dagger)^{j'},
\ee
with $U$ being a generic $N\times N$ unitary matrix and $\bar{U}$ its conjugate. So far, the species are decoupled, so we couple them by the insertion of a boundary term at $x=0$
\be\label{eq:S_bound}
S_{\text{boundary}} = \int d\tau (\bar{\Psi}^\dagger)^j(x=0,\tau)S_{jj'}\Psi^{j'}(x=0,\tau) + (\Psi^\dagger)^j(x=0,\tau)S^{\dagger}_{jj'}\bar{\Psi}^{j'}(x=0,\tau),
\ee
where $S$ is a unitary $N\times N$ matrix parametrizing the mixing. Notice that, while in general the $U(N)$ symmetry is broken, the $U(1)$ symmetry associated with the imbalance between particle and quasi-particle is conserved. A naive way to characterize the unbroken group $H \subset U(N)$, is to identify the set of unitary transformations which leave the boundary term \eqref{eq:S_bound} invariant. We thus require that $U \in H$ iff
\be
\bar{U}_{lj}S_{lm}U_{mj'} = S_{jj'},
\ee
where repeated indices are summed over. Being $U$ unitary, we can rephrase the condition above as $U^{-1}SU=S$ or, equivalently, $[S,U]=0$.

To establish the validity of the previous considerations, we aim to characterize the string order parameter systematically via BCFT methods. We start by identifying the boundary state $\ket{b}$ of our model as \cite{cmc-22} 
\be
\ket{b}=\prod_{k\in \mathbb{N}-\frac{1}{2}}\exp\left(iS_{jj^{'}}\Psi^{\dag j}_{-k}\bar\Psi^{j'}_{-k}+(\Psi\to\Psi^{\dag})\right)\ket{0}.
\ee
The total Hamiltonian is just given by the sum of the single-specie Hamiltonian, and it is
\be
L_0+\bar L_0 = \sum_{k>0}k \left(\Psi^j_{-k}\Psi^j_{k} + \bar{\Psi}^j_{-k}\bar{\Psi}^j_{k} + (\Psi \rightarrow \Psi^\dagger )\right).
\ee
The last key ingredient is the action of the symmetry on the fermionic modes. We associate to any $U \in U(N)$ an operator $\hat{U}$ satisfying
\be\label{eq:Un_commrel}
\begin{split}
\hat{U}\Psi^j_{-k}=U_{j'j}\Psi^{j'}_{-k}\hat{U}, \quad \hat{U}\bar{\Psi}^j_{-k}=U_{j'j}\bar{\Psi}^{j'}_{-k}\hat{U},\\
\hat{U}(\Psi^\dagger)^j_{-k}=\bar{U}_{j'j}(\Psi^\dagger)^{j'}_{-k}\hat{U}, \quad \hat{U}(\bar{\Psi}^\dagger)^j_{-k}=\bar{U}_{j'j}(\bar{\Psi}^\dagger)^{j'}_{-k}\hat{U},
\end{split}
\ee
which is equivalent to Eq. \eqref{eq:UN_symmetry}. Putting everything together, we compute the charged partition function
\be
\begin{split}
&\mathcal{Z}(U) \equiv \bra{b}q^{L_0+\bar{L}_0}\hat{U}\ket{b} = \\
&\prod_{k>0}\bra{0}\exp(-iS^{\dag}_{jj'^{'}}\bar{\Psi}^j_{k} (\Psi^\dagger)_{k}^{j^{'}})\hat{U}q^{L_0+\bar L_0}\exp(iS_{jj^{'}} (\Psi^\dagger)^j_{-k}\bar{\Psi}^{j^{'}}_{-k})\ket{0} \times (\Psi \rightarrow \Psi^\dagger).
\end{split}
\ee
Using the commutation relations \eqref{eq:Un_commrel} and the invariance of the vacuum $\ket{0}$ under the $U(N)$ symmetry, we get
\be\label{chargedfermi}
\begin{split}
&\bra{0}\exp(-iS^{\dag}_{jj'^{'}}\bar{\Psi}^j_{k} (\Psi^\dagger)_{k}^{j^{'}})\hat{U}q^{L_0+\bar L_0}\exp(iS_{jj^{'}} (\Psi^\dagger)^j_{-k}\bar{\Psi}^{j^{'}}_{-k})\ket{0} =\\
&\bra{0}\exp(-iS^{\dag}_{jj'^{'}}\bar{\Psi}^j_{k} (\Psi^\dagger)_{k}^{j^{'}})\exp(iq^{2k}(U^\dagger SU)_{jj^{'}} (\Psi^\dagger)^j_{-k}\bar{\Psi}^{j^{'}}_{-k})\ket{0} =\\
&\det \left(1+q^{2k}S^{\dag}U^{\dag}SU\right),
\end{split}
\ee
where in the final step we applied the formula
\be
\bra{0}\exp\left(\mathcal {O^{\prime}}_{jj^{\prime}}\Psi_{k}^j\bar\Psi_{k}^{j^{\prime}}\right)
\exp\left(\mathcal {O^{\prime}}_{jj^{\prime}}\Psi_{-k}^j\bar\Psi_{-k}^{j^{\prime}}\right)\ket{0}=\det\left(1-\mathcal{O^{\prime}}\mathcal{O}\right)\,\,,
\ee
proved in \cite{cmc-22}. Summing over the modes $k$, we finally reach to
\be
\mathcal Z(U) = \prod_{k\in \mathbb{N}-1/2} \left|\det\left( 1+q^{2k} S^{\dagger}U^{\dagger} S U \right)\right|^2.
\ee
To proceed further with the computation, it is convenient to introduce the $N\times N$ unitary matrix
\be\label{eq:O_matrix}
\mathcal O\equiv S^{\dag}U^{\dag}SU.
\ee
In this way, we express
\be
\begin{split}
\det (1+q^{2k}\mathcal{O})&=\prod_{\lambda \in \text{Spec}(\mathcal{O})}(1+q^{2k}\lambda),\\
\det (1+q^{2k}\mathcal{O}^\dagger)&=\prod_{\lambda \in \text{Spec}(\mathcal{O})}(1+q^{2k}\lambda^{-1}),
\end{split}
\ee
with $\text{Spec}(\mathcal{O})$ being the set of eigenvalues of $\mathcal{O}$. After a bit of algebra, we finally reach an exact expression for the string-order parameter in the large $L/\varepsilon$ limit
\be
-\log \bra{\Omega}\hat{U}\ket{\Omega} = -\log \frac{\mathcal{Z}(U)}{\mathcal{Z}(1)} = \frac{1}{4\pi^2} \log \frac{L}{\varepsilon} \sum_{\lambda \in \text{Spec}(\mathcal{O})}(\text{Li}_2(-\l)+\text{Li}_2(-\l^{-1})-2\text{Li}_2(-1)).
\ee
At this point, we want to understand for which $U\in U(N)$ the order parameter vanishes, providing a characterization of the unbroken group $H$. We first observe that for $|\lambda|=1$ it holds
\be
\text{Li}_2(-\l)+\text{Li}_2(-\l^{-1})-2\text{Li}_2(-1)\geq 0,
\ee
and the inequality is saturated only for $\lambda=1$. This implies that $-\log \bra{\Omega}\hat{U}\ket{\Omega} =0$ iff every eigenvalue of $\mathcal{O}$ is $1$. In other words, the order parameter vanishes when $\mathcal{O} =1$, a condition equivalent to (see Eq. \eqref{eq:O_matrix})
\be\label{eq:commute}
[S,U] =0.
\ee
This is not particularly surprising, as the naive argument based on the invariance of the boundary action leads to the same conclusion. Nevertheless, it establishes its validity and allows us to identify the unbroken group as
\be
H = \{U \in U(N) \ | \ [U,S] =0 \}.
\ee

\section{Complex bosons}
\label{sec:Bosons}

In this section, we generalize the same symmetry breaking patterns discussed  for fermions in Sec. \ref{sec:Fermions} to free complex massless bosons\footnote{For more details in the calculations of this section the reader may consult \cite{cmc2-22}.}. Although many analogies can be recognized, and the underlying physics is similar, the analytical predictions for the order parameters differ explicitly.

\subsection{$U(1)$-symmetry breaking terms in complex bosons}

\begin{figure}[ht]
\centering
	\includegraphics[width=0.9\linewidth]{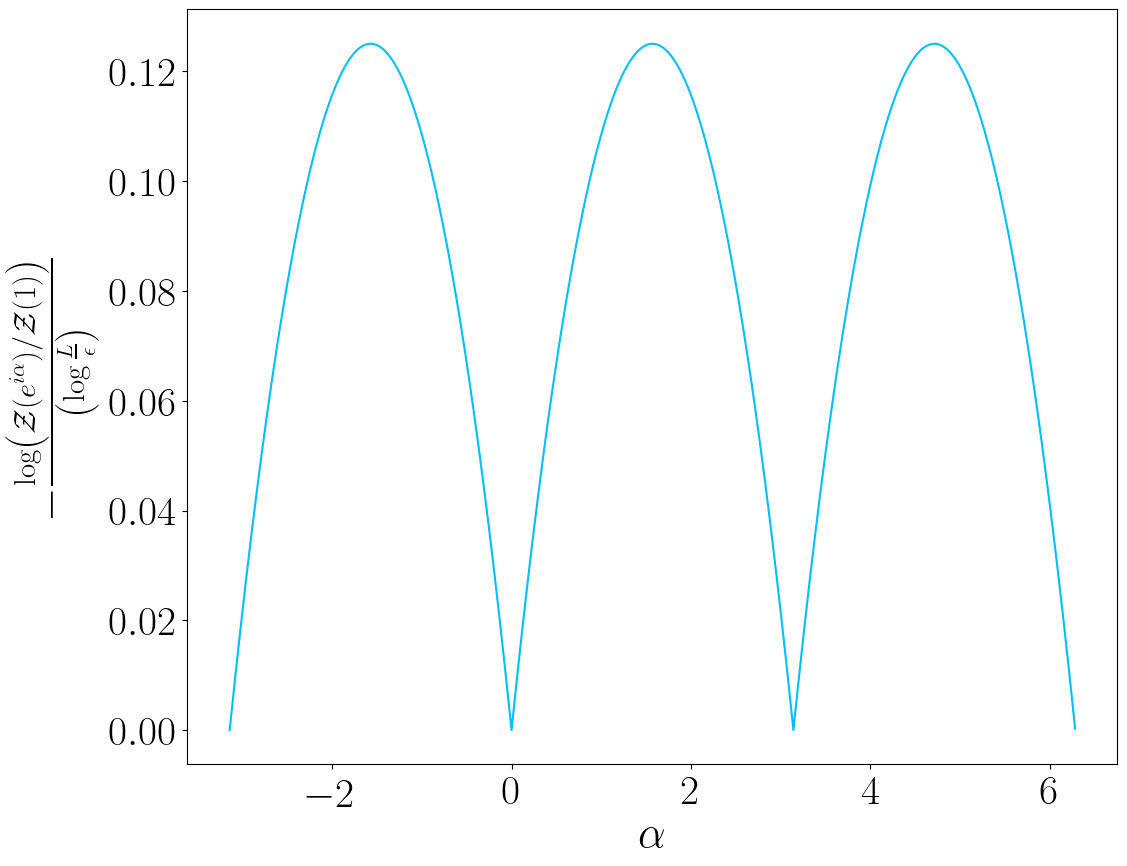}
    \caption{String order parameter as function of $\alpha$ for the complex boson. The function is periodic for $\alpha \rightarrow \alpha + \pi$, and shows cusp singularities in correspondence of $\alpha= k\pi, k\in \mathbb{Z}$. It vanishes at $\alpha= k\pi, k\in \mathbb{Z}$ and it signals the presence of unbroken $\mathbb{Z}_2$ subgroup.
    }
    \label{fig:SOPbosons}
\end{figure}

First, we consider the $U(1)$-symmetry action
\be
S_{\text{bulk}}=\int_0^Ldx\int d\tau\,\del\Phi\,\bar\del\Phi^{\dag}.
\ee
We expand the bosonic field in its Laurent modes as
\be
\Phi(z)=\sum_{k\in \mathbb{Z}}\,\frac{\Phi_k}{z^{k}}\,\,,\qquad \bar \Phi(\bar z)=\sum_{k\in \mathbb{Z}}\,\frac{\bar\Phi_k}{\bar z^{k}}\,\,.\\
\ee
The bulk action is invariant under the $U(1)$ symmetry
\be\label{U1bosons}
(\Phi,\bar\Phi)\to e^{i\a}(\Phi,\bar\Phi)\,\,,\qquad (\Phi^{\dag},\bar\Phi^{\dag})\to e^{-i\a}(\Phi^{\dag},\bar\Phi^{\dag}).
\ee
We are interested in a boundary breaking term which breaks $U(1)$ explicitly and preserves a $\mathbb{Z}_2$ symmetry, as in \eqref{U1breaking}. In analogy with the fermionic case, we take
\be
S_{\text{boundary}}=\int d\tau\, \left(\Phi(x=0,\tau)\bar\del\Phi(x=0,\tau) +(\Phi\rightarrow \Phi^\dagger)\right)\,.
\ee
Our aim is to compute the string-order parameter via BCFT. To abridge words, using the experience from fermions, we need the boundary states, corresponding to the chosen BCs, the symmetry generator and the Hamiltonian of the system. These are given respectively by 
\be
|b\rangle=\prod_{k>0}\exp\left(\Phi_{-k}\bar\Phi_{-k}+
(\Phi\to\Phi^{\dag})\right)|0\rangle,
\ee
\be
Q=\sum_{k>0}\left(\Phi_{-k}\Phi_k+\bar\Phi_{-k}\bar\Phi_k-(\Phi\to\Phi^{\dag})\right),
\ee
and 
\be
L_0+\bar L_0=\sum_{k>0}k\left(\Phi_{-k}\Phi_k+\bar\Phi_{-k}\bar\Phi_k+(\Phi\to\Phi^{\dag})\right).
\ee
In terms of these quantities, the partition function can be expressed by \eqref{partition}, as for fermions, with $q$ given by \eqref{eq:def_q}. We now proceed with the evaluation, decomposing the partition function as a product of bosonic modes
\be
\begin{split}
\mathcal Z(e^{i\a})=&\prod_{k>0}\langle0|\exp\left(\bar\Phi_{k}\Phi_{k}\right)\exp\left(i\a \Phi_{-k}\Phi_k+i\a \bar\Phi_{-k}\bar\Phi_k\right)\\
&\times q^{k\Phi_{-k}\Phi_{k}+k\bar\Phi_{-k}\bar\Phi_k}\exp\left(\Phi_{-k}\bar\Phi_{-k}\right)|0\rangle\times \left(\Phi\to \Phi^{\dag}\right).
\end{split}
\ee
The contribution from the $k$-th mode of $\Phi$ is given by
\be
\begin{split}
\langle0|\exp\left(\bar\Phi_{k}\Phi_{k}\right)\exp\left(i\a \Phi_{-k}\Phi_k+i\a \bar\Phi_{-k}\bar\Phi_k\right) q^{k\Phi_{-k}\Phi_{k}+k\bar\Phi_{-k}\bar\Phi_k}\exp\left(\Phi_{-k}\bar\Phi_{-k}\right)|0\rangle=\\
\bra{0}\exp(\bar{\Phi}_{k} \Phi_{k})\exp(q^{2k}e^{i2\alpha} \Phi_{-k}\bar\Phi_{-k})\ket{0}  = (1-q^{2k}e^{i2\alpha})^{-1},
\end{split}
\ee
where we used the commutation relations and the properties  $\Phi_k\ket{0}=0$ and $\bar\Phi_k\ket{0}=0$, for $k>0$. Taking the contribution of $\Phi^{\dag}$ and putting everything together, we obtain

\be
\mathcal Z(e^{i\a})=\prod_{k>0}\left(1-q^{2k}e^{2i\a}\right)^{-1}\left(1-q^{2k}e^{-2i\a}\right)^{-1}=\prod^{\infty}_{m=1}\left(1-2\cos(2\a)q^{2m}+q^{4m}\right)^{-1}\,.
\ee
We notice that the property
\be
\mathcal Z(-1)=\mathcal Z(1),
\ee
holds also here, so that a $\mathbb{Z}_2$ symmetry is unbroken. The $L/\e\to \infty$ limit is obtained by converting the infinite product to an integral 
\be
\begin{split}
\log \mathcal Z(e^{i\a})&\simeq-\int_0^{\infty}dk\log\left(1-q^{2k}e^{i2\a}\right)+\log\left(1-q^{2k}e^{-i2\a}\right)\\
&=-\frac{1}{2\log q}\left(\text{Li}_2(e^{i2\alpha})+\text{Li}_2(e^{-i2\alpha})\right).
\end{split}
\ee
In the limit above, we express the order parameter as
\be
-\log\bra{\Omega}e^{i\a Q}\ket{\Omega}=-\log\frac{\mathcal Z(e^{i\a})}{\mathcal Z(1)}=\log \frac{L}{\varepsilon}\left( \frac{\alpha}{2\pi}-\frac{\alpha^2}{2\pi^2} \right), \quad \alpha \in [0,\pi],
\ee
and its value is periodic under $\alpha \rightarrow \alpha+\pi$. Finally, it is worth to recognize explicitly the presence of cusp singularities at $\alpha=0$ and $\alpha=\pi$, which were absent in the fermionic counterpart. We show the $\alpha$-dependence of the order parameter in Fig. \ref{fig:SOPbosons}.

\subsection{$U(N)$-symmetry breaking via boundary scattering matrix}
We now consider the bosonic version of the model in \ref{sec:UN_ferm}. The bulk action is 
\be
S_{\text{bulk}}=\int_0^Ldx\int d\tau\,\del\Phi^j\,\bar\del\Phi^{\dag j}.
\ee
where the sum over the specie index $j$ ($j=1,\dots,N$) is implicit. The action is invariant under the following $U(N)$ transformation
\be\label{eq:UN_bosons}
\Phi^j \rightarrow U_{j'j}\Phi^{j'}, \quad  \bar{\Phi}^j \rightarrow U_{j'j}\bar{\Phi}^{j'}, \quad  (\Phi^\dagger)^j \rightarrow \bar{U}_{j'j}(\Phi^\dagger)^{j'}, \quad 
(\bar{\Phi}^\dagger)^j \rightarrow \bar{U}_{j'j}(\bar{\Phi}^\dagger)^{j'}.
\ee
We take the following boundary action
\be
S_\text{boundary}=\int d\tau \left((\bar\Phi^{\dag})^j(0,\tau)S^{\dagger}_{jj^{'}}\Phi^{j^{'}}(0,\tau)+(\Phi^{\dag})^j(0,\tau)S_{jj^{'}}(\bar\Phi)^{j^{'}}(0,\tau)\right),
\ee
where $S$ is a $N\times N$ matrix satisfying similar properties with the fermionic case.

The boundary state, the charge operator and the total Hamiltonian are respectively 
\be
\begin{split}
\ket b&=\prod_{k>0}\exp\left( S_{jj^{'}}\Phi^{\dag j}_{-k}\bar\Phi^{j^{'}}_{-k}+\Phi\to\Phi^{\dag}\right),\\
Q&=\sum_{k>0}\left(\Phi_{-k}^j\Phi_k^j+\bar\Phi^j_{-k}\bar\Phi_k^j-(\Phi\to\Phi^{\dag})\right),
\end{split}
\ee
\be
L_0+\bar L_0=\sum_{k>0}k\left(\Phi_{-k}^j\Phi_k^j+\bar\Phi^j_{-k}\bar\Phi_k^j+(\Phi\to\Phi^{\dag})\right).
\ee
The charged partition function reads
\be\label{chargedboson}
\begin{split}
&\mathcal Z(U)\equiv\bra b \hat U q^{L_0+\bar L_0}\ket b=\\
&\prod_{k>0}\bra0\exp\left( S^{\dag}_{jj^{'}}\bar\Phi_k^j(\Phi^{\dag})^{j^{'}}_k\right)\hat Uq^{L_0+\bar L_0}\exp\left(S_{jj^{'}}(\Phi^{\dag})^j_{-k}\bar\Phi^{j^{'}}_{-k}\right)\ket0\times\left(\Phi\to\Phi^{\dag}\right)\\
&=\prod_{k>0}|\det(1-q^{2k}S^{\dag}U^{\dag}SU)|^{-2},
\end{split}
\ee
where in the last line we applied the formula
\be
\bra{0}\exp\left(\mathcal {O^{\prime}}_{jj^{\prime}}\Phi_{k}^j\bar\Phi_{k}^{j^{\prime}}\right)
\exp\left(\mathcal {O^{\prime}}_{jj^{\prime}}\Phi_{-k}^j\bar\Phi_{-k}^{j^{\prime}}\right)\ket{0}=\det\left(1-\mathcal{O^{\prime}}\mathcal{O}\right)^{-1},\,\,
\ee
proven in  \cite{cmc2-22}. Using the definition \eqref{eq:O_matrix} and the relations
\be
\begin{split}
   \det(1-q^{2k}\mathcal O)^{-1}&=\prod_{{\l\in Spec(\mathcal O)}}(1-q^{2k}\l)^{-1}, \\
   \det(1-q^{2k}\mathcal O^\dag)^{-1}&=\prod_{{\l\in Spec(\mathcal O)}}(1-q^{2k}\l^{-1})^{-1},
\end{split}
\ee
one finally obtains the expression of the string order parameter in the large $L/\epsilon$ limit
\be
-\log\bra\Omega U\ket \Omega=-\log\frac{\mathcal Z(U)}{\mathcal Z(1)}=\frac{1}{4\p^2}\log\frac{L}{\e}\sum_{\l\in Spec(\mathcal O)}\left(\text{Li}_2(\l)+\text{Li}_2(\l^{-1})-2\text{Li}_2(1)\right).
\ee
Here, as for fermions, one observes that the order parameter vanishes exactly for $\mathcal{O}=1$, which means $[S,U]=0$.

\section{Lattice results}
\label{sec:Lattice}

In this section, we provide some lattice realizations of free fermions with a boundary symmetry breaking, whose scaling regime is captured by the BCFTs described above.\\
For instance, while we are not aware of any lattice realization of the field theory described in Sec.\ref{sec:U1_ferm}, characterized by the boundary-breaking of a $U(1)$ symmetry, we consider a system that we conjecture to be its doubling. We describe that system, relating its properties to those of the homogeneous counterpart. Finally, we evaluate numerically the order parameter in the lattice, and we compare it to the analytical predictions of Sec. \ref{sec:U1_ferm}.

Then, we consider a Fermi chain with the insertion of a conformal defect, whose scattering properties do not explicitly depend on the incoming momentum \cite{ep-12,ep2-12,ce-22}. This system can be regarded as a theory of two species of particles on the half-line coupled together at a boundary point, via the so-called unfolding procedure, and its underlying  BCFT is described \ref{sec:UN_ferm}. In particular, the $U(2)$ symmetry associated with the mixing of the two species is broken explicitly due to BCs, encoded in the scattering matrix of the defect.

\subsection{$U(1)$-symmetry breaking by a boundary pairing: doubling trick}

Let us first consider the homogeneous hamiltonian
\be
\label{eq:hopping}
\mathcal{H}=-\sum_{x}[c_x^{\dagger }c_{x+1}+h.c.],
\ee
describing free fermions hopping on the lattice. Here $c_x^\dagger, c_x$ are fermionic creation and annihilation operators associated with the site $x$, verifying anticommutation relations
\be
\{c_x^{\dagger},c_{x'}\}=\delta_{xx'}, \quad \{c_x,c_{x'}\}=0, \quad \{c_x^{\dagger},c_{x'}^{\dagger}\}=0.
\ee
We now describe the previous hamiltonian in terms of a new set of fermionic operators $a_x$ and $a_x^{\dagger}$, defined by
\be 
\label{eq:map}
c_x=\begin{cases} a_x, & \mbox{if } x\leq 0 \\ (-1)^xa_x^{\dagger}, & \mbox{if } x >0,\end{cases}
\ee
that amounts to the exchange of the role of creation and annihilation operators in the right half of the system. The explicit expression of the hamiltonian after the mapping becomes
\be 
\label{eq:Ham2}
\mathcal{H}=-\left( \sum_{x\leq 0}(a_x^{\dagger} a_{x+1}+h.c.)+a_0a_1+a_1^{\dagger}a_0^{\dagger}+\sum_{x \geq 1}(a_x^{\dagger}a_{x+1}+h.c.) \right).
\ee
In these new variables, $\mathcal{H}$ is no longer homogeneous, and a pairing term appears as a localized defect between the sites $x=0,1$. Moreover, the $U(1)$ symmetry
\be
a_x \rightarrow a_x e^{i\theta}, \quad \quad a^\dagger_x \rightarrow a^\dagger_x e^{-i\theta},
\ee
shared by the bulk terms of the hamiltonian, is broken explicitly due to the defect.
Before analyzing further the lattice system, we provide a heuristic argument to explain the relationship between the Hamiltonian \eqref{eq:Ham2} and the CFT in Sec \ref{sec:U1_ferm}. Let us consider the vacuum\footnote{This is not the ground state of $\mathcal{H}$. The latter will be denoted below with $\ket{\Omega}$.} $\ket{0}$ as the state satisfying
\be
c_x\ket{0} = 0, \quad  \forall x.
\ee
One can consider one-particle excitations of $\ket{0}$, generated by linear combinations of $\{c^\dagger_x\}$ acting on the vacuum $\ket{0}$. Since in the formulation  Eq. \eqref{eq:hopping} the system is homogeneous, an incoming wave-packet would reach the point $x=0$ and propagate across it without being partially reflected. The same process can be described in the language of the fermions $\{a_x\}_x$  given the formulation \eqref{eq:Ham2}. We first notice that $\ket{0}$ satisfies
\be
a_x |0\rangle =0  \quad \mbox{if}\quad x>0, \hspace{2 cm} a^{\dagger}_x |0\rangle =0  \quad \mbox{if}\quad x\leq 0,
\ee
namely that the left/right chain is completely empty/filled. An incoming right-moving excitation can be thus interpreted as a particle that hits the central point, is completely transmitted, and then becomes a hole. Similarly, if we considered a Fermi sea, the excitations would have been given by particles/holes which change their $U(1)$ charge after the scattering at $x=0$. This mechanism is nothing but an explicit symmetry breaking, and its origin can be traced in the term $a_0a_1+a_1^{\dagger}a_0^{\dagger}$ of \eqref{eq:Ham2}. A similar scenario has been depicted in Sec. \ref{sec:U1_ferm}. The crucial difference is that, while in the lattice system \eqref{eq:Ham2} one can recognize both left and right-moving incoming particles, the incoming particles of \ref{sec:U1_ferm} are just left-moving. Heuristically, we thus conjecture that \eqref{eq:Ham2} is a discretization of the QFT depicted above once the degrees of freedom are doubled.

To better motivate this argument, in the following we analyze the order parameter associated with the $U(1)$-symmetry breaking. We firstly regularize the hamiltonian $\mathcal{H}$ in \eqref{eq:hopping}, keeping the size of the system finite $x\in[-L+1,L]$, and we consider its ground state $\ket{\Omega}$. In the language of the fermionic operators $c_x$, it can be regarded as a Fermi sea at half-filling, namely, the total number of particles is $N=L$, and it satisfies
\be 
\left(\sum_{x\leq 0} c_x^{\dagger}c_x+\sum_{x>0}c_x^{\dagger}c_x\right)|\Omega \rangle =N|\Omega \rangle.
\ee
While the hamiltonian is not changed after the mapping \eqref{eq:map}, and so its ground state $\ket{\Omega}$, the description in terms of the operators $a_x$ is less transparent, as particles and holes (antiparticles) are mixed among each other by the pairing. A preliminary observation is that
\be
\left(\sum_{x\leq 0} a_x^{\dagger}a_x-\sum_{x>0}a_x^{\dagger}a_x\right)|\Omega \rangle = N|\Omega\rangle,
\ee
namely, the imbalance of particles among the left/right half-chain is fixed, and it does not fluctuate. Then, we express the generator of the $U(1)$ symmetry  $a_x\rightarrow a_x e^{i\theta}$ as
\be 
Q=\sum_{x}a_x^{\dagger}a_x=\sum_{x\leq 0} c_x^{\dagger}c_x-\sum_{x>0}c_x^{\dagger}c_x = N_L-N_R,
\ee
with $N_{L,R}$ the number operator in the left/right chain before the mapping \eqref{eq:map}. Crucially, the ground state fluctuations of $Q$ are strictly related to those of $2N_L$, as it holds
\be
Q\ket{\Omega} = (N_L-N_R)\ket{\Omega} = (2N_L-N)\ket{\Omega},
\ee
and $N=L$ is just an additive constant. In this way, we can finally express the full-counting statistics of $Q$ as 
\be
\label{eq:FCSmap}
\bra{\Omega}e^{i\alpha Q}\ket{\Omega}  \sim\bra{\Omega}e^{i2\alpha N_L}\ket{\Omega},
\ee
up to an irrelevant proportionality constant. In conclusion, Eq. \eqref{eq:FCSmap} gives an effective way to characterize the order parameter of the $U(1)$ breaking in Eq. \eqref{eq:Ham2} to the full counting statistics of a subsystem in the homogeneous chain \eqref{eq:hopping}, which is easier to compute.

We now show how to express $\bra{\Omega}e^{i2\alpha N_L}\ket{\Omega}$, employing standard techniques for free fermions \cite{ep-09}. We first construct the correlation matrix of the ground state, defined as
\be
C_{xx'}  \equiv \bra{\Omega}c^\dagger_x c_{x'} \ket{\Omega}.
\ee
Since $\ket{\Omega}$ is a Fermi sea with $N=L$ particles, one can express
\be
C_{xx'} = \sum^{L}_{j=1}\overline{\phi_j(x)}\phi_j(x'), \quad x \in [-L+1,L]
\ee
with $\phi_j(x)$ being the single-particle eigenfunction of the Hamiltonian \eqref{eq:hopping}. Then, we restrict the spatial indices to the left side of the chain, obtaining a $L\times L$ matrix $C_A$ satisfying
\be
(C_A)_{xx'} = C_{xx'}, \quad x \in [-L+1,0].
\ee
Given the eigenvalues of $C_A$, denoted by $\text{Spec}(C_A) = \{\nu_j\}_{j=1,\dots,L}$, one can show that the following relation holds
\be
\langle e^{i2\alpha N_L }\rangle =\prod^{L}_{j=1}[\nu_j e^{2i\alpha}+(1-\nu_j)].
\ee
Making use of Eq. \eqref{eq:FCSmap}, one finally gets the string-order parameter as
\be 
\label{eq:FCSnum}
\bra{\Omega} e^{i\alpha Q}\ket{\Omega} =\langle e^{i2\alpha N_L }\rangle =\prod^{L}_{j=1}[\nu_j e^{2i\alpha}+(1-\nu_j)].
\ee
An explicit analytical expression for the eigenvalues $\{\nu_j\}_j$ is a hard task, and should rely on numerics. Nevertheless, simple field theoretical arguments can capture the correct leading behavior of the order parameter, as we explain below. The state $\ket{\Omega}$ can be recovered as a euclidean path integral of the massless Dirac fermions over the strip
\be
x \in [-L,L], \quad \tau  \in (-\infty,\infty),
\ee
with $\tau$ representing the Euclidean time, and we assume $L$ to be much larger than the lattice size. Furthermore, the operator $e^{i\alpha N_L}$ is described by the insertion of two vertex operators in the path integral at $\tau=0$, as
\be
e^{i\alpha N_L} \sim \mathcal{V}_\alpha(x=-L)\mathcal{V}_{-\alpha}(x=0).
\ee
It is known \cite{ch-05} that the bulk scaling dimension $\Delta_\alpha$ of $\mathcal{V}_\alpha$ is
\be
\Delta_{\alpha} = \left(\frac{\alpha}{2\pi}\right)^2, \quad \alpha \in [-\pi,\pi].
\ee
Moreover, the insertion of the boundary operator $\mathcal{V}_\alpha(x=-L)$ does not play a role, since the BCs at $x=-L$ are $U(1)$ symmetric, and from now on we just discard it. Making use of scale-invariance, we finally reach to
\be
\bra{\Omega}e^{i2\alpha N_L}\ket{\Omega} \sim \bra{\Omega}\mathcal{V}_{-2\alpha}(x=0)\ket{\Omega} \sim \frac{1}{L^{\Delta_{2\alpha}}} = \frac{1}{L^{\alpha^2/\pi^2}}, \quad \alpha \in [-\pi/2,\pi/2],
\ee
where we employed scaling arguments on the bulk field. In conclusion, thanks to \eqref{eq:FCSmap}, we finally obtain
\be\label{eq:OrdPar_FermLat}
-\log |\bra{\Omega}e^{i\alpha Q}\ket{\Omega} |  \simeq \frac{\alpha^2}{\pi^2}\log L , \quad \alpha \in [-\pi/2,\pi/2]
\ee
which is the main result of this section. We emphasize that the formula we obtained is just twice the prediction in Section \ref{sec:U1_ferm}. This is somehow expected, as we conjectured that the Hamiltonian \eqref{eq:Ham2} describes a doubling of the system in Sec. \ref{sec:U1_ferm}.

\begin{figure}[t]
\centering
	\includegraphics[width=0.9\linewidth]{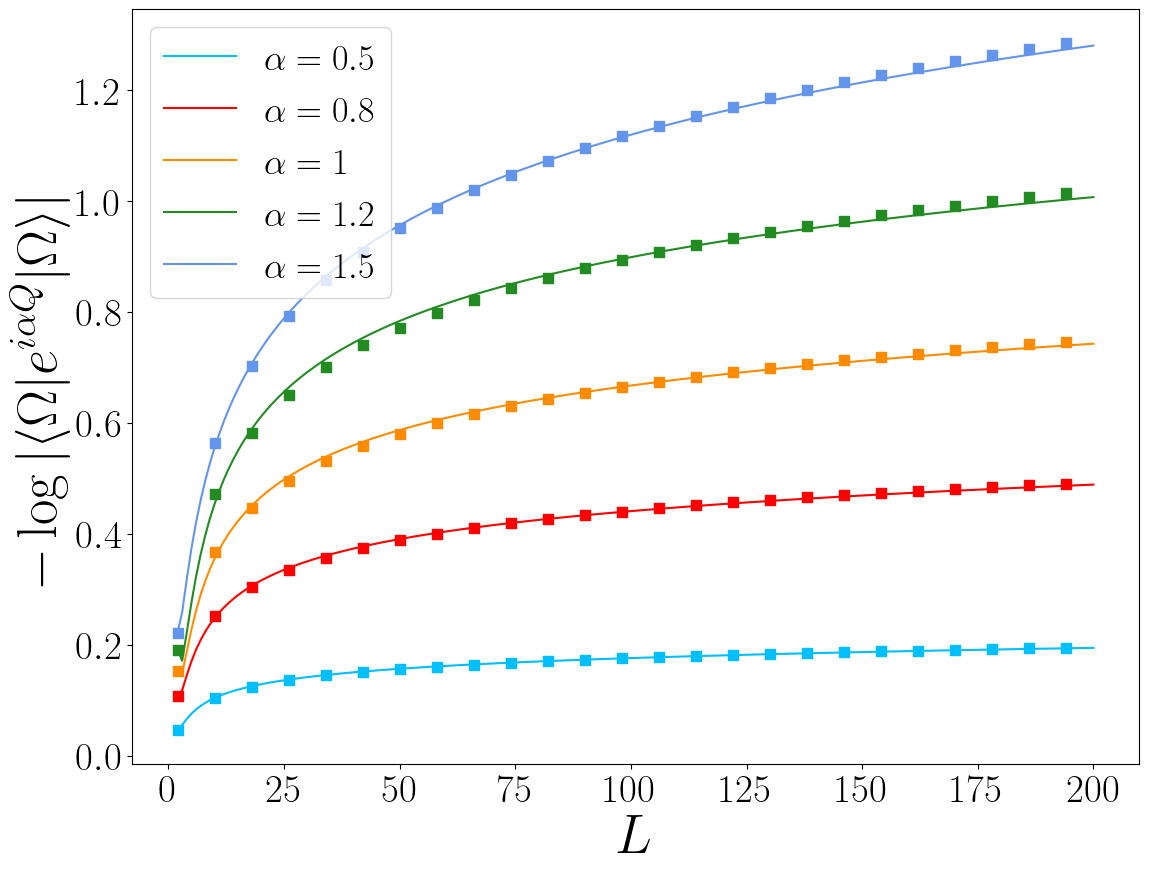}
    \caption{ Order parameter as function of $L$ for different values of $\alpha$. The symbols are the numerical data obtained from Eq.\eqref{eq:FCSnum}. The solid lines show the curve $(\alpha^2/\pi^2)\log L + b_0 + b_1L^{-1}$, where the first term is the analytical prediction given in Eq. \eqref{eq:OrdPar_FermLat} and coefficients $b_0,b_1$ are obtained fitting the data.  
    }
    \label{fig:FCSfermions}
\end{figure}

\subsection{Conformal defect and $U(2)$ symmetry breaking}

Let us consider a system made of two species of spinless fermions on the lattice $x\in [0,L-1]$, coupled together at the boundary point $x = 0$ through a conformal defect. 

The hamiltonian is
\be
\mathcal{H}= -\sum_{x,j }[(c_x^{\dagger})^jc_{x+1}^j+h.c.] - \sum_{jj'} S_{jj'} (c^\dagger_0)^j(c_0)^{j'},
\ee
where $(c_x^\dagger)^j, (c_x)^j$ are fermionic operators verifying anticommutation relations
\be
\{(c_x^{\dagger})^j,(c_{x'})^{j'}\}=\delta_{xx'}\delta_{jj'}, \quad \{(c_x)^j,(c_{x'})^{j'}\}=0, \quad \{(c_x^{\dagger})^j,(c_{x'}^{\dagger})^{j'}\}=0, \quad j=1,2,
\ee
and $S$ is the following $2\times2$ matrix
\be
S=\begin{pmatrix}
\sqrt{1-\lambda^2}&\lambda\\
\lambda &-\sqrt{1-\lambda^2}
\end{pmatrix}, \quad \lambda \in [0,1].
\ee
It is possible to show that the scattering matrix, induced by the boundary term in the hamiltonian, is exactly $S$ \cite{bddo-02,bm-06,ce-22}. Since the transmission/reflection probability does not depend on the incoming momenta of the particle, and it is given by $\lambda^2,1-\lambda^2$ respectively, the defect is scale-invariant, and it is described by a  BCFT. It is easy to show that the bulk terms on the hamiltonian are invariant under a $U(2)$ symmetry which mixes the two species as
\be
c_x^j \rightarrow U_{j'j}c_x^{j'}, \quad \quad (c_x^{\dagger})^j \rightarrow \bar{U}_{j'j} (c_x^{\dagger})^{j'},
\ee
with $U$ a generic 2x2 unitary matrix, and the sum over $j'$ is implicit. However, the presence of the defect breaks explicitly the symmetry. Indeed, after a generic $U(2)$ transformation, the boundary term is mapped onto
\be
-S_{jj'}\overline{U}_{j_1 j}U_{j_2 j'} (c^\dagger_0)^{j_1}(c_0)^{j_2},
\ee
and it is invariant only if $U^\dagger S U = S$, namely $[U,S] =0$. While the condition above was already derived in CFT, it is instructive to notice that it holds in the lattice model too.\\
For the sake of completeness, we characterize explicitly the unbroken subgroup $H$ defined by
\be 
H=\left\{U\in U(2)| [S,H] =0 \right\}.
\ee
Since $S$ is hermitian and unitary, its eigenvalues have to be real phases, and they are just $\pm 1$. In the basis in which $S$ is diagonal, also $U \in H$ has to be diagonal, since $S$ and $H$ commute, and so they share the same eigenspaces. This means that, given a matrix $D$ which diagonalizes $S$, say
\be
S=D\begin{pmatrix}
1&0\\
0 &-1
\end{pmatrix}D^{-1}, \quad \quad D=\begin{pmatrix}
\frac{\lambda}{\sqrt{2-2\sqrt{1-\lambda^2}}}&\frac{\lambda}{\sqrt{2+2\sqrt{1-\lambda^2}}}\\
\frac{\sqrt{1-\sqrt{1-\lambda^2}}}{\sqrt{2}} & -\frac{\sqrt{1+\sqrt{1-\lambda^2}}}{\sqrt{2}} 
\end{pmatrix},
\ee
we identify the unbroken group as
\be 
H=\left\{U=D\begin{pmatrix}
e^{i \alpha_1}&0\\
0 &e^{i \alpha_2}
\end{pmatrix}D^{-1}, \alpha_1, \alpha_2 \in \mathbb{R}\right\}.
\ee

\section{Conclusions and Outlooks}\label{sec:Conclusions}

In this work, we considered the effect of an explicit symmetry breaking in a one-dimensional critical system induced by a localized impurity. We develop a general formalism to probe the symmetry breaking that can be applied to any system described by a BCFT. In particular, we show that $-\log |\bra{\Omega}\hat{g}\ket{\Omega}|$, which is a non-local order parameter, is generically logarithmic growing in the system size, and it vanishes precisely for the unbroken elements of the group. We provide exact calculations for free theories, fermions and bosons, with both abelian and non-abelian broken symmetries. For instance, we first consider the symmetry breaking of a complex theory in the presence of a boundary pairing term, which has a residual $\mathbb{Z}_2$ symmetry. Then, we consider a theory made of many species of particles coupled together at a boundary point, and a non-trivial breaking of a $U(N)$ symmetry is investigated.

A natural generalization would be the calculation of the order parameter for rational minimal models (Ising, Potts model, Tricritical Ising, and so on)\cite{DiFrancesco-97}. Indeed, their scale-invariant BCs are known, and they are mapped onto the so-called Cardy states. The evaluation of the overlap in Eq. \eqref{eq:charged_pfun} might be more involved for interacting CFTs, but we think it should be expressed as Virasoro characters, whose explicit expressions are known.

One may also wonder what happens when the theory is not described by a BCFT, or due to a finite correlation length in the bulk $\xi$ or in the presence of BCs which are not scale invariant  \cite{c-84,s-88,gz-94,lcmss-95,klcm-96,c-04,bpt-06,cad-09}. 
In the first case, the boundary effect is localized in a typical distance $\xi$ and we expect that the order parameter goes to a constant when the system size is increased. It might be interesting to study this scenario in Integrable Field Theories, providing exact results for massive theories.
In the second case, the BCs are expected to flow to some scale-invariant ones via RG flow. Then, for a big enough size, the leading growth of the order parameter would be logarithmic, with a universal prefactor depending only on the IR fixed point. We hope to come back to these problems in the future.

Finally, we mention that while our approach involves a genuine non-local probe of the system, one might be interested in the local observables. In particular, it has been proposed \cite{amc-22} that a subsystem measure dubbed as \text{entanglement-asymmetry} might be a good probe of the symmetry breaking. An interesting direction would be the computation of the entanglement asymmetry for an interval attached to the boundary point in BCFT.

\newpage

\section*{Acknowledgments}

RB~acknowledges support from the Croatian Science Foundation (HrZZ) project No. IP-2019-4-3321. LC  acknowledges support from ERC
under Consolidator grant number 771536 (NEMO). The  work of P. Panopoulos was supported by the Croatian Science Foundation Project "New Geometries for Gravity and Spacetime" (IP-2018-01-7615).

\end{document}